\input{aipcheck}

\documentclass[
    ,final            
  ]
  {aipproc}

\layoutstyle{8x11double}

\newcommand{\beq}{\begin{equation}}
\newcommand{\eeq}{\end{equation}}

\newcommand{\bea}{\begin{eqnarray}}
\newcommand{\eea}{\end{eqnarray}}
\newcommand{\ba}{\begin{array}}
\newcommand{\ea}{\end{array}}

\begin{document}

\title{How to check that dyons are at work?}


\author{Dmitri Diakonov}
{
address={Petersburg Nuclear Physics Institute, Gatchina 188300, St. Petersburg, Russia},
altaddress={Ruhr-Universit\"at Bochum, Bochum 44780, Germany}
}

\begin{abstract}
I present arguments in favor of the dyon mechanism of confinement and deconfinement. Dyons may be those
real physical objects that are revealed as lattice monopoles in the Abelian gauges, and as central
vortices in the central gauges. I suggest a lattice calculation of the effective action as function
of the gauge invariant eigenvalues of the Polyakov line, which is of general interest but in particular
may support or refute the importance of dyons in the Yang--Mills vacuum.

\end{abstract}

\maketitle


There is multiple evidence from theory, lattice and phenomenology that dyons play
an important role in enacting confinement, and in inducing deconfinement phase transition:
\begin{itemize}
\item In ${\cal N}=1$ supersymmetric theory it is precisely dyons that shape the vacuum, in particular the gluino condensate,
and it is an {\em exact} result there~\cite{Davies:1999uw,Davies:2000nw,Diakonov:2002qw}
\item Lattice studies show that zero fermion modes `jump' from one position in space to another as one varies
fermion boundary conditions -- precisely as one would expect from dyons carrying zero modes~\cite{Gattringer:2003uq}
\item a semiclassical picture of the vacuum populated by dyons gives an appealing explanation of all main features
associated with confinement: the area behavior of large Wilson loops, the asymptotic linear rising potential only for
nonzero $N$-ality probes, the cancelation of gluons in the free energy, and a 1$^{\rm st}$ order deconfinement
phase transition for all gauge groups except $SU(2)$, regardless of whether the group has a nontrivial center or not~\cite{Diakonov:2007nv,Diakonov:2010qg}.
\end{itemize}

In a few words, if the Polyakov line is, on the average, not an element of the group center, as in the confinement
phase, dyons appear as saddle points of the Yang--Mills (YM) partition function. Dyons are gluon field configurations
with asymptotic Coulomb-like chromo-electric and -magnetic fields. The ensemble of many dyons is similar to a
multi-component plasma. In particular, the dual (magnetic) gluons obtain a Debye mass. This is the physical reason
for the exponential decay of the Polyakov lines correlations, {\it i.e.} of the linear rising potential for heavy
probe `quarks'. The appearance of a mass for dual gluons is also responsible for the area behavior of large Wilson
loops. A surface spanning the loop is a source for a soliton of finite thickness, `made' of the dual fields.
The action per area of this soliton is the string tension which coincides, at low temperatures $T$,
with the string tension computed from the correlation function of the Polyakov lines~\cite{Diakonov:2007nv}.

The ensemble of dyons induce a nonperturbative energy that is a function of the Polyakov loop eigenvalues.
The function is such that its minimum, for any gauge group, is at a universal set of eigenvalues related
to the Weyl vector, the half-sum of positive roots of the Lie algebra for the gauge group~\cite{Diakonov:2010qg}.
For most gauge groups it implies that the trace of the Polyakov loop is zero in the lowest dimensional
representations (however there are subtleties related to the fluctuations about the minimum). This dyon-induced
nonperturbative potential energy competes with the well-known perturbative potential energy, also a function
of the Polyakov loop eigenvalues. The perturbative energy scales as $T^4$ with respect to the nonperturbative one.
Therefore, at some critical $T_c$ it prevails, and that is the mechanism for the deconfinement phase transition.
It happens irrespectively of what is the center of the gauge group. \\

There are presently several qualitative pictures of confinement being discussed. Apart from dyons, these
are Abelian monopoles and center vortices, see {\it e.g.}~\cite{Greensite:2009zz}. It may be that
all three pictures can be, in a sense, reconciled. Dyons may be the real physical objects that reveal
themselves as Abelian monopoles seen on the lattice in the Abelian gauges, such as the maximal Abelian gauge,
and also as center vortices observed in the center gauges, such as the maximal center gauge.

Indeed, in order to assemble many dyons together they need to have the same asymptotic field $A_4$ at spatial
infinity. That necessarily requires that dyons are in the `stringy' gauge where a singular Dirac string is
sticking out from each dyon. The Dirac strings are gauge artifacts: the action density is finite there. However,
they do carry a quantized Abelian magnetic flux. When, in a lattice simulation, one uses any variant of the
Abelian gauge, one identifies lattice magnetic monopoles as the sources of that flux. The exact position of
Abelian lattice monopoles varies somewhat as one varies gauge fixing -- in accordance with the fact that the
direction of the Dirac string sticking from a dyon is subject to a gauge choice. However, lattice magnetic monopoles
may well be a reflection of real physical objects, the dyons. It would be interesting to check it directly.

Furthermore, if one further restricts the gauge to a center gauge, center vortices are
revealed~\cite{Greensite:2009zz}. They can be understood as the Dirac strings connecting dyons.
There are gauges where a dyon has a Dirac string entering it, and another leaving it.
See a recent study in Ref.~\cite{Bruckmann:2010bs} of the relation between dyons and vortices.

I should also mention a talk by Langfeld and Ilgenfritz at this conference~\cite{Langfeld:2010ue}, 
who ``cooled'' lattice configurations keeping Polyakov loops fixed. Usually smearing the configurations by cooling
kills confinement but in this study it is preserved. The interesting observation is that the
``cooled'' configurations preserving confinement are mainly (anti)self-dual fields.
Instantons and dyons are (anti)self-dual.

Concerning instantons, the quantum Coulomb interactions of dyons are such that they tend to glue up into
electric- and magnetic-neutral clusters which at low temperatures are hardly distinguishable from
instantons~\cite{Diakonov:2009jq}. The difference with the old `instanton liquid' model~\cite{Diakonov:1983hh}
is that ({\it i}) the Polyakov line is now nontrivial, ({\it ii}) the integration measure over collective
coordinates is invariant under permutation of dyons `belonging' to different instantons, and allows instantons
to overlap. These circumstances are critical for obtaining confinement that was absent in former instanton
models.\\

Finally, I would like to point out a (simple) lattice measurement that may help to understand
the nature of the YM vacuum in general, and to demonstrate (or refute) the importance of dyons, in particular.
I suggest to measure the effective action for the gauge-invariant eigenvalues of the Polyakov line. The definition
is given in Eq.(3) of Ref.~\cite{Diakonov:2010qg}. In lattice setting, one puts all time links to be unity matrices
(the $A_4=0$ gauge), but make the spatial links periodic up to a gauge transformation with the matrix $L({\bf x})$
being the Polyakov line. Without loss of generality one can take it to be diagonal, as in Eq.(1) of
Ref.~\cite{Diakonov:2010qg}. One then simulates the ensemble of configurations with fixed `eigenphases'
$\mbox{\boldmath$\phi$}({\bf x})$ as boundary conditions. In particular, one can take $\mbox{\boldmath$\phi$}$
to be ${\bf x}$-independent. The partition function or the free energy itself is not calculable by Monte Carlo
methods but one can find the average plaquette and then integrate it over $\beta$ or temperature to obtain
the free energy. This will be the effective potential as function of $\mbox{\boldmath$\phi$}$.
At large $T$ it is the well-known perturbative potential energy as function of $\mbox{\boldmath$\phi$}$.
It is interesting to see how it looks like below and above $T_c$ for different groups.

If, as we assume, dyons are of relevance, the minimum of the effective potential will be at a specific point
$\mbox{\boldmath$\phi$}$ proportional to the Weyl vector $\mbox{\boldmath$\rho$}$, for any gauge group,
see Eq.(5) of Ref.~\cite{Diakonov:2010qg}.

It should be noted that if one just studies the distribution of the Polyakov line `eigenphases', it will be
in any case dominated by the Haar measure weight that governs the local ultraviolet quantum fluctuations.
In order to see the smooth potential energy as function of $\mbox{\boldmath$\phi$}$, one really needs to consider
the case of a constant or slowly varying Polyakov lines using, for example, the setting described above.\\

I thank Christof Gattringer and Victor Petrov for helpful discussions. Mercator Professorship
by Deutsche Forschungsgemeinschaft (DFG), and a partial support by the Russian
Foundation for Basic Research, grant 09-02-01198, are gratefully acknowledged.

\bibliographystyle{aipproc}   

\begin{thebibliography}{99}

\bibitem{Davies:1999uw}
  N.~M.~Davies, T.~J.~Hollowood, V.~V.~Khoze and M.~P.~Mattis,
  Nucl.\ Phys.\  B {\bf 559}, 123 (1999)

\bibitem{Davies:2000nw}
  N.~M.~Davies, T.~J.~Hollowood and V.~V.~Khoze,
  J.\ Math.\ Phys.\  {\bf 44}, 3640 (2003)

\bibitem{Diakonov:2002qw}
  D.~Diakonov and V.~Petrov,
  Phys.\ Rev.\  D {\bf 67}, 105007 (2003)

\bibitem{Gattringer:2003uq}
  C.~Gattringer {\it et al.},
  Nucl.\ Phys.\ Proc.\ Suppl.\  {\bf 129}, 653 (2004)

\bibitem{Diakonov:2007nv}
D.~Diakonov and V.~Petrov, Phys.\ Rev.\ D {\bf 76}, 056001 (2007)

\bibitem{Diakonov:2010qg}
  D.~Diakonov and V.~Petrov, these Proceedings,
  arXiv:1011.5636 [hep-th].

\bibitem{Greensite:2009zz}
  J.~Greensite,
  Acta Phys.\ Polon.\  B {\bf 40}, 3355 (2009)

\bibitem{Bruckmann:2010bs}
  F.~Bruckmann, E.~M.~Ilgenfritz, B.~Martemyanov and B.~Zhang,
  these Proceedings, arXiv:1011.6178 [hep-th];
  Phys.\ Rev.\  D {\bf 81}, 074501 (2010)

\bibitem{Langfeld:2010ue}
  K.~Langfeld and E.~M.~Ilgenfritz, these Proceedings, 
  arXiv:1012.2308 [hep-lat].
also
  arXiv:1012.1214 [hep-lat];
  K.~Langfeld
  arXiv:0911.0319 [hep-lat].

\bibitem{Diakonov:2009jq}
D.~Diakonov, Nucl. Phys. Proc. Suppl. {\bf 195}, 5 (2009)

\bibitem{Diakonov:1983hh}
  D.~Diakonov and V.~Y.~Petrov,
  Nucl.\ Phys.\  B {\bf 245}, 259 (1984);
  Nucl.\ Phys.\  B {\bf 272}, 457 (1986).

\end{thebibliography}

\end{document}